\documentclass[12pt,preprint]{emulateapj}

\usepackage{graphicx}
\usepackage{amsmath}

\begin{document}

\title{Measuring the distances to quasars at high redshifts with strong lensing}
\author{Kai Liao$^{1}$}
\affil{
$^1$ {School of Science, Wuhan University of Technology, Wuhan 430070, China.}}
\email{liaokai@whut.edu.cn}

\begin{abstract}
Strongly lensed quasars with time-delay measurements are well known to provide
the ``time-delay distances" $D_{\Delta t}=(1+z_L)D_LD_S/D_{LS}$
and the angular diameter distances to lens galaxies $D_L$. These two kinds of distances
give stringent constraints on cosmological parameters. In this work, we explore a different use of time-delay observables:
Under the assumption of a flat Universe, strong lensing observations can accurately measure
the angular diameter distances to sources $D_S$. The corresponding redshifts of
quasars may be up to $z_S\sim4$ according to the forecast.
The high-redshift distances would
sample the Hubble diagram between SNe Ia and CMB, cosmological-model-independently providing direct information on the evolution of the nature of our Universe, for example,
the dark energy Equation-of-State parameter $w(z)$.
We apply our method to the existing lensing system SDSS 1206+4332 and get $D_S=2388_{-978}^{+2632}Mpc$ at $z_S=1.789$. We
also make a forecast for the era of LSST. The uncertainty of $D_S$ depends on the redshifts of lens and source, the
uncertainties of $D_{\Delta t}$ and $D_L$, and the correlation between $D_{\Delta t}$ and $D_L$ as well. Larger correlation would result in tighter $D_S$ determination.

\end{abstract}

\section{Introduction}
In the standard cosmological model, i.e., $\Lambda$CDM, the Universe is flat, dominated by cold dark matter and dark energy
with Equation-of-State (EoS) parameter $w\equiv-1$~\citep{Frieman2008}. This concordance scenario is able to explain most of the cosmological observations. However,
more and more issues have emerged~\citep{Moore1994,Moore1999,Frieman2008,Ding2015}. Especially, the Hubble constant ($H_0$) measurement based on Cepheids and Type Ia
Supernovae (SNe) from local Universe~\citep{Riess2019,Freedman2017}
has $4.4\sigma$ discrepancy with measurement from Cosmic Microwave Background (CMB) that assumes the $\Lambda$CDM model
when inferring $H_0$~\citep{Planck2018}.
A recent independent determination of the $H_0$ based on the Tip of the Red Giant Branch (TRGB)
seems to reduce the discrepancy~\citep{Freedman2019}.
The inconsistency problem would either be related with unknown systematic errors or reveal new physics beyond the standard model.
Alterative cosmological models were proposed to solve these issues while new problems may be brought in.

Putting aside the models, from the observational perspective, it is crucial to reconstruct the expansion history of the Universe directly and
model-independently from the data~\citep{Shafieloo2006,Shafieloo2007,Clarkson2010}.
Cosmology-free calculations can be also seen in~\citep{Bernal2016,Li2019,Arendse2019,Denissenya2018}.
If we have distances measured at different redshifts $z$, we can reconstruct the distance-redshift relation $D(z)$, i.e.,
the Hubble diagram using cosmology-independent methods,
for example, the Gaussian process~\citep{Shafieloo2012,Seikel2012,Yang2015}.
In addition, we can also reconstruct other cosmological quantities evolving with redshift, for example, the Hubble expansion rate $H(z)$, the deceleration parameter $q(z)$ and
the dark energy EoS parameter $w(z)$. These would in turn help us understand the issues related with theoretical models.
However, the reconstruction is limited by the maximum redshift $z_{max}$ of the data~\citep{LHuillier2019}.
Note that current reliable data used to study cosmology are either at low redshifts $z<2$, for example SNe Ia~\citep{Betoule2014} and the
Baryon Acoustic Oscillations (BAO)~\citep{Anselmi2019}, or very high redshift $z\sim1000$,
i. e., the CMB~\citep{Planck2018}.
Other cosmological approaches, for example,
cosmic chronometers~\citep{Chen2017} and galaxy clusters~\citep{Chen2012} are also at low redshfits.
Therefore, it is important to get high-redshift data
(hereafter we take $z>2$ as ``high-redshift") to fill up the data desert between the farthest SN Ia and CMB.
The Gamma Ray Bursts (GRBs) can be observed up to $z\sim8$ and may provide the distance measurements~\citep{Schaefer2003,Izzo2009,Wei2010}. However, they need
calibration by SNe Ia at low redshifts which is very uncertain. Issues also exist about the physical motivation of GRBs as standard candles~\citep{Wang2015}.
Gravitational waves by compact binary stars can also provide luminosity distances as standard sirens~\citep{Schutz1986}, however, the measurement uncertainty
would increase remarkably at redshift $z>2$~\citep{Cai2017,Zhao2018}. Recently, quasars at redshifts up
to $z\sim5$ were proposed to measure the luminosity distances with a method based on
X-ray and ultraviolet emission~\citep{Risaliti2019}. The robustness of this method needs to be further confirmed.

Strongly lensed quasars by galaxies are an excellent tool to study astrophysics and cosmology~\citep{Treu2010}.
The distant AGN with its host galaxy is lensed by the foreground elliptical galaxy, forming multiple
images and arcs of the host galaxy. With the measurements of time-delays between these images, the
``time-delay distance" which is a combination of three angular diameter distances $D_{\Delta t}=(1+z_L)D_LD_S/D_{LS}$ can be determined~\citep{Refsdal1964,Treu2016}.
It is known to determine the Hubble constant~\citep{Refsdal1964} and
other cosmological parameters, for example, the EoS parameter of dark energy~\citep{Linder2011}.
The H0LiCOW collaboration~\citep{Suyu2017} has constrained $H_0$ at $2.4\%$ precision level under a flat $\Lambda$CDM model
and a weaker constraint in $w$CDM model due to the degeneracy between $H_0$ and $w$~\citep{Wong2019,Taubenberger2019}.
In addition to $D_{\Delta t}$, the angular diameter distance to lens galaxy ($D_L$) was proposed to be determined by combining
time-delay measurements with lens stellar velocity dispersion measurements~\citep{Paraficz2009,Jee2015,Jee2016,Yildirim2019}.
Four lensing systems have been given the robust $D_L$ measurements~\citep{Wong2019}.
Note that unlike SNe Ia which determine the relative distances, the strong lensing measures the absolute angular diameter distances,
with which one can directly establish the Hubble diagram.
However, this approach is limited by the relatively low redshifts of the lenses $z<1.2$~\citep{Jee2016}.

Motivated by acquiring high-redshift data for studying the Universe,
we propose a method to measure the distances to quasars based on strong lensing. The source quasars locate at
high redshifts up to $z_S\sim4$. This paper is organised as follows: In Section 2, we introduce the current status of time-delay strong lensing cosmology.
In Section 3, we give the idea of measuring the distances to the sources. Then we apply our method to
a realistic system SDSS 1206+4332 in Section 4. We also make a forecast for the lensing observations in LSST era in Section 5.
Finally, we summarize and make discussions in Section 6.

\section{Lensed quasars with time-delays}
According to the theory of strong gravitational lensing~\citep{Refsdal1964,Treu2010,Treu2016,Liao2019}, the arriving time difference (time-delay) between two images of the source measured
from Active Galactic Nucleus (AGN) light curves
is related with the geometry of the Universe and the gravity field of lens galaxy through:
\begin{equation}
\Delta t=\frac{D_{\Delta t}}{c}\Delta \phi(\boldsymbol{\xi}_{lens}),\label{Dt}
\end{equation}
where c is the light speed.
$\Delta\phi=[(\boldsymbol{\theta}_A-\boldsymbol{\beta})^2/2-\psi(\boldsymbol{\theta}_A)-(\boldsymbol{\theta}_B-\boldsymbol{\beta})^2/2+\psi(\boldsymbol{\theta}_B)]$
is the Fermat potential difference between image A and image B.
$\boldsymbol{\theta}_A$ and $\boldsymbol{\theta}_B$ are angular positions of the images.
$\boldsymbol{\beta}$ denotes the source angular position.
$\psi$ is the two-dimensional lensing potential determined by the Poisson equation $\nabla^2\psi=2\kappa$,
where the surface mass density of the lens
$\kappa$ is in units of critical density $\Sigma_{\mathrm{crit}}=c^2D_S/(4\pi GD_LD_{LS})$.
$\Delta \phi$ is determined by the lens model parameters $\boldsymbol{\xi}_{lens}$ which can be inferred with high resolution imaging data.
$D_{\Delta t}$ is the ``time-delay distance" consisting of three angular diameter distances:
\begin{equation}
D_{\Delta t}=(1+z_L)\frac{D_LD_S}{D_{LS}},\label{defDt}
\end{equation}
where $L,S$ stands for lens and source.
Note that the line-of-sight (LOS) mass structure could also affect time-delay distance measurements~\citep{Falco1985,Rusu2017}.

At the same time, the angular diameter distance ratio can be measured in a general
form (not limited to a Singular Isothermal Sphere (SIS) model as one usually takes):
\begin{equation}
\frac{D_{LS}}{D_S}=\frac{c^2J(\boldsymbol{\xi}_{lens},\boldsymbol{\xi}_{light},\beta_{ani})}{(\sigma^P)^2},\label{ratio}
\end{equation}
where $\sigma^P$ is the LOS projected stellar velocity dispersion of the lens galaxy which provides extra constraints to the cosmographic inference.
The parameter $J$ captures all the model components computed from angles measured on the sky (the imaging) and the stellar orbital anisotropy distribution.
It can be written as a function of lens model parameters $\boldsymbol{\xi}_{lens}$,
the light profile parameters $\boldsymbol{\xi}_{light}$ and the anisotropy distribution of the
stellar orbits $\beta_{ani}$.

Combining Eq.\ref{Dt} and Eq.\ref{ratio}, the angular diameter distance to the lens can be measured by:
\begin{equation}
D_L=\frac{1}{1+z_L}\frac{c\Delta t}{\Delta \phi(\boldsymbol{\xi}_{lens})}\frac{c^2J(\boldsymbol{\xi}_{lens},\boldsymbol{\xi}_{light},\beta_{ani})}{(\sigma^P)^2}.
\end{equation}
Note that the lensing analysis is quite complicated and we only show the key equations. For dealing with the real data,
one should use a full Bayesian analysis considering covariances between quantities to calculate the posteriors of each parameter.
We refer to Jee et al. (2015), Shajib et al. (2018), Birrer et al. (2019) and Y${\i}$ld${\i}$r${\i}$m et al. (2019) for more details of such process.

Therefore, the lensed quasars with time-delays could constrain parameters in cosmological models through the measured $D_{\Delta t}$ and $D_L$.
The H0LiCOW project~\citep{Suyu2017} in collaboration with the COSMOGRAIL programme~\citep{Courbin2018} has assembled a sample of lensed quasar systems, six of which (B1608+656, RXJ1131-1231, HE 0435-1223, SDSS 1206+4332, WFI2033-4723, PG 1115+080) have been well-analyzed in the milestone paper~\citep{Wong2019}.
Among them, four systems have both $D_{\Delta t}$ and $D_L$ measurements.
Currently, the H0LiCOW team has only published the posteriors of both $D_{\Delta t}$ and $D_L$ measurements for SDSS 1206+4332.
Assuming a flat $\Lambda$CDM and through a blind analysis, they reported $H_0=73.3_{-1.8}^{+1.7}km/s/Mpc$, a $2.4\%$ precision including systematics.
A detailed results in different models can be found in~\citep{Wong2019}. The previous results from only 4 systems can be found in~\citep{Taubenberger2019}.

\section{Distances to the sources}

We propose in this work that strong lensing can also provide the angular diameter distances to quasars ($D_S$). As long as one assumes the Universe is flat,
the three relevant angular diameter distances can be respectively expressed as:

\begin{equation}
D_L=\frac{c}{(1+z_L)H_0}\int_0^{z_L}\frac{1}{E(z)}dz,
\end{equation}

\begin{equation}
D_S=\frac{c}{(1+z_S)H_0}\int_0^{z_S}\frac{1}{E(z)}dz,
\end{equation}
and
\begin{equation}
D_{LS}=\frac{c}{(1+z_S)H_0}\int_{z_L}^{z_S}\frac{1}{E(z)}dz=D_S-\frac{1+z_L}{1+z_S}D_L,
\end{equation}
where $E(z)=H(z)/H_0$. Therefore, with Eq.\ref{defDt}, the $D_S$ can be determined by:
\begin{equation}
D_S=\frac{(1+z_L)D_LD_{\Delta t}}{(1+z_S)\left[D_{\Delta t}-(1+z_L)D_L\right]}.\label{Ds}
\end{equation}
In other words, if lensing observations give $D_{\Delta t}$ and $D_L$, one can always measure (infer)
$D_S$ equivalently. We emphasize that although determining $D_S$ in this way would not bring any benefits (extra information) for constraining
parameters in specific cosmological models, the measured $D_S$ at high-redshifts can be further used in
the model-independent reconstruction of the expansion history of the Universe, whereas $D_L$ can be replaced by other low-redshift data, for example, the SNe Ia.
When applying Eq.\ref{Ds}, one should consider the correlation between $D_{\Delta t}$ and $D_L$. Actually, from strong lensing observations, only one distance among $D_L$, $D_S$ and $D_{\Delta t}$ is totally independent. While the community uses either $D_L$ or $D_{\Delta t}$, we focus on $D_S$ in this work.
In principle, rather than inferring it from $D_{\Delta t}$ and $D_L$,
one can directly take $D_S$ as the lensing parameter in the first place during the lensing analysis.
For example, for the simplest case where the lens is describe by a SIS model~\citep{Paraficz2009}, the density distribution is given by:
\begin{equation}
\rho_{SIS}(r)=\frac{\sigma^2}{2\pi Gr^2},
\end{equation}
where $\sigma^2$ is the three-dimensional isotropic velocity dispersion. Then
\begin{equation}
D_{\Delta t}=\frac{2c\Delta t}{\theta^2_A-\theta^2_B},\label{Dtsis}
\end{equation}
and
\begin{equation}
D_L=\frac{c^3\Delta t}{4\pi\sigma^2(1+z_L)\Delta\theta}.\label{DLsis}
\end{equation}
Thus we can directly relate $D_S$ with observations by combining Eq.\ref{Ds}, Eq.\ref{Dtsis} and Eq.\ref{DLsis}.

\begin{figure}
 \includegraphics[width=8cm,angle=0]{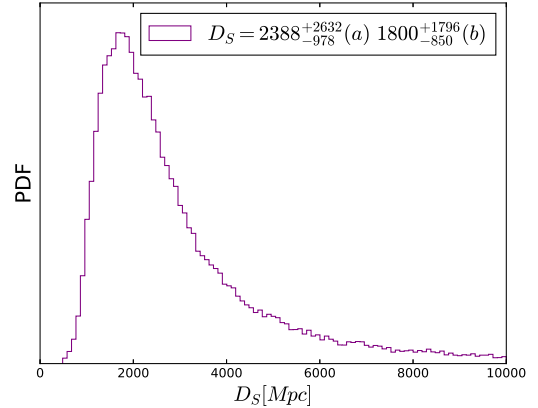}
  \caption{Measurement of the distance to the source for SDSS 1206+4332. We adopt two statistics for the distribution:
   (a) The mean value plus the $16^{th}$ and $84^{th}$ percentiles; (b) The most probable value plus $68\%$ probability. The lower and upper
   limits have the same probability density.
  } \label{result}
\end{figure}

\begin{figure}
 \includegraphics[width=8cm,angle=0]{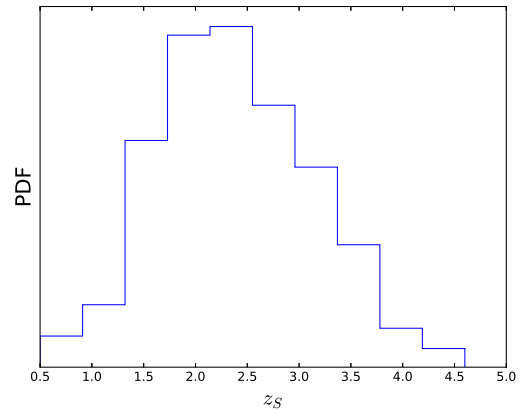}
  \caption{The redshift distribution of the sources. 65$\%$ of the systems would have $z_S>2$ resulting 35 high-redshift distance measurements.
  } \label{zs}
\end{figure}

While we take the SIS model for illustration purposes, one should note that realistic lenses are much more
complicated. Different components in the mass models were explored both for lensing and for kinematics~\citep{Jiang2007,Jee2015,Shajib2018a,Shajib2018}.
For the macro mass model, one needs to consider more properties for individual lenses, for example, the ellipticity and the density slope.
A singular elliptical power-law model may not be sufficient and one usually tries a composite model consisting of a baryonic component linked
to the stellar light distribution plus an elliptical NFW dark matter halo. In some cases, the lens is during a merger process, for example, B1608+656
shows two interacting lens galaxies~\citep{Suyu2010}. Besides, substructures, for example, the satellites and the dark matter sub-halos would
make observations anomalous if one ignore them~\citep{Liao2018}. Furthermore, the nearby galaxies, the line-of-sight structure can also make the lens modelling complicated.

\section{Measurement of SDSS 1206+4332}

\begin{figure*}
 \includegraphics[width=16cm,angle=0]{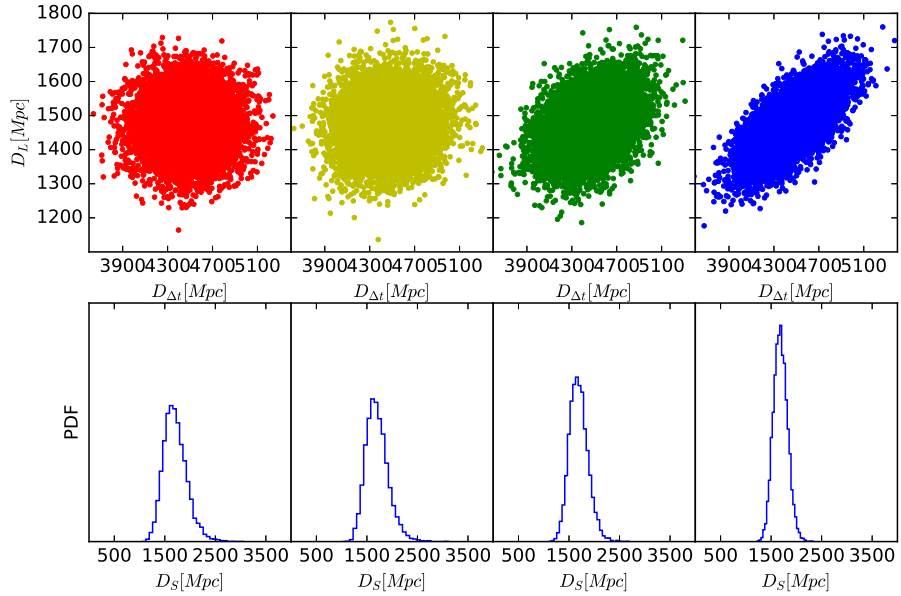}
  \caption{A typical case with $z_L=0.7$ and $z_S=2.5$. The uncertainties of $D_{\Delta t}$ and $D_L$ are set by $5\%$.
  The upper panels show the simulated $D_{\Delta t}$ and $D_L$ distributions with different correlation amplitudes:
  $\rho=0, 0.1, 0.4, 0.7$, respectively. The bottom panels are the corresponding $D_S$ inferences.
  } \label{example1}
\end{figure*}

\begin{figure*}
 \includegraphics[width=16cm,angle=0]{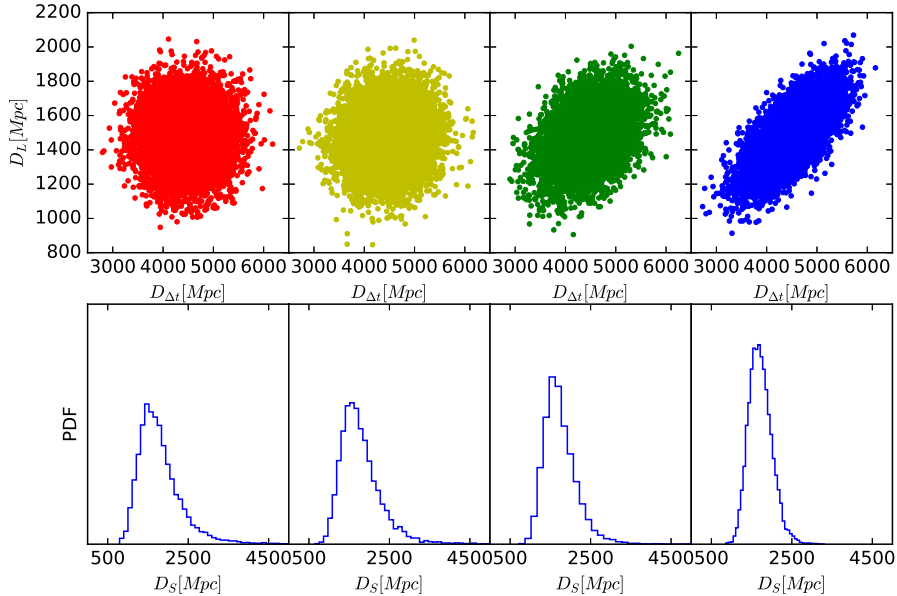}
  \caption{The same as Fig.\ref{example1} but for $10\%$ uncertainties of $D_{\Delta t}$ and $D_L$.
  }\label{example2}
\end{figure*}

\begin{figure*}
 \includegraphics[width=16cm,angle=0]{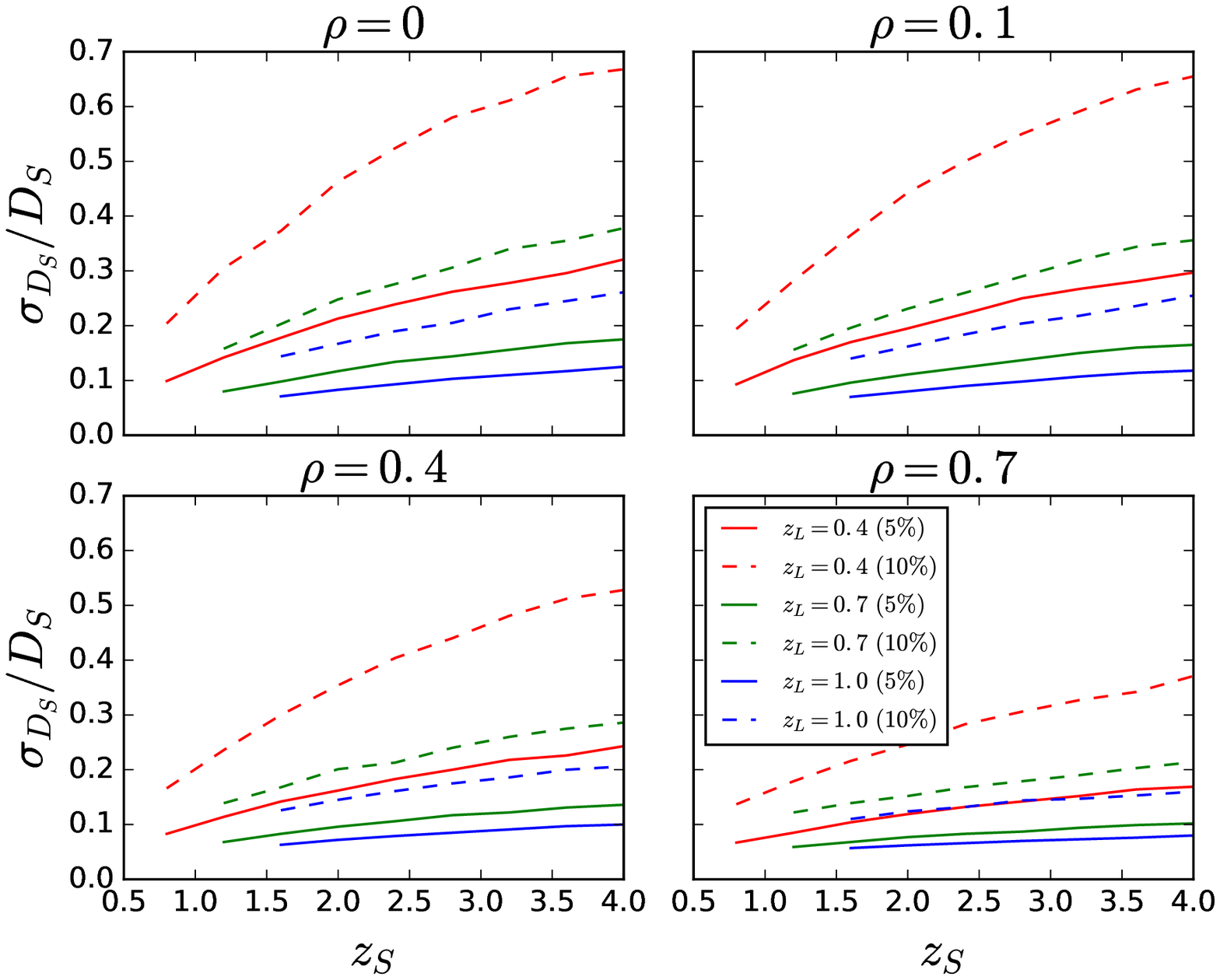}
  \caption{The relative uncertainties of the $D_S$ measurements for different lens redshift $z_L$ and source redshift $z_S$. The precisions of
  $D_{\Delta t}$ and $D_L$ are taken as $5\%$ and $10\%$ respectively. The impacts of different correlation amplitudes between $D_{\Delta t}$ and $D_L$
  are shown in the 4 subfigures, respectively.
  } \label{forecast}
\end{figure*}

We apply our method to the system SDSS 1206+4332 which was discovered by~\citep{Oguri2005}.
It is one of high-quality lensing systesms in the catalog of H0LiCOW and has been modeled by~\citep{Birrer2019} within the program.
This system consists of a doubly lensed quasar with the host galaxy forming a nearly-complete Einstein ring.
The image separation is $3.03''$ and the time-delay measured from the light curve pair is $111.3\pm3$ days. The redshifts of lens and
source are $z_L=0.745$ and $z_S=1.789$, respectively.
The H0LiCOW team took a blind time-delay strong lensing cosmographic analysis of the system. They combined the time-delay measurement between
the two AGN images, Hubble Space Telescope imaging, spectroscopic data of the lens galaxy and the line-of-sight field measurements to give measurements of both
$D_L$ and $D_{\Delta t}$ with systematic errors under control.
The two distances
are provided on the website\footnote{http://www.h0licow.org} of the program in the form of tables of parameters sampled from the posterior
distributions. Note the correlation between them is quite slight with current uncertainties.

For each pair of ($D_{\Delta t}$, $D_L$) in the posterior tables, i.e., each point in the two dimensional marginalized distributions of $D_{\Delta t}$ and $D_L$
(see Fig.12 in Birrer et al. 2019),
we calculate $D_S$ based on Eq.\ref{Ds} considering the correlation in this way. The distribution of $D_S$ is shown in Fig.\ref{result}.
The median value plus the $16^{th}$ and $84^{th}$ percentiles is $D_S=2388_{-978}^{+2632}Mpc$.
We notice the distribution is very deviated from Guassian-like, therefore, we also give the most
probable value $D_S=1800_{-850}^{+1796}Mpc$, where the lower and upper limits have the same
probability density and include $68\%$ probability.
We approximately fit the distribution in Fig.\ref{result} with a log-normal function:
\begin{equation}
\small
P(D_{\Delta t})=\frac{1}{\sqrt{2\pi}(x-\lambda_D)\sigma_D}exp\left[-\frac{(ln(x-\lambda_D)-\mu_D)^2}{2\sigma_D^2}\right],
\end{equation}
where $x=D_{\Delta t}/(1 Mpc)$ and the parameters $(\lambda_D=1050,\sigma_D=1.10,\mu_D=7.43)$.

Note that SDSS 1206+080 is the only doubly lensed system among the current 6 H0LiCOW lenses~\citep{Wong2019}, $D_{\Delta t}=5769_{-471}^{+589}Mpc$ and $D_L=1805_{-398}^{+555}Mpc$.
The uncertainties of $D_{\Delta t}$ and $D_L$ are $\sim9\%$ and $\sim26\%$. For quadruply lensed systems, the uncertainties would
be much smaller. For example, $D_{\Delta t}$ is constrained with $\sim4.3\%$ precision in RXJ1131-12131, and 
$D_L$ is constrained with $\sim13\%$ in B1608+656~\citep{Wong2019}.

\section{Forecast in the LSST era}
An increasing number of lensed quasars are being discovered by the current surveys, for example, the Dark Energy Survey (DES)
and the Hyper Suprime-Cam Survey (HSC). Moreover, the upcoming Large Synoptic Survey Telescope (LSST)~\citep{OM10}
will bring us thousands of lensed quasars, some of which will have
long-time high-quality light curves for each lensed image. The Time Delay Challenge (TDC) program~\citep{Liao2015} has proved that
with good algorithms, there will be $\sim400$
well-measured time-delays with average precision $\sim3\%$ and the average bias $<1\%$.
According to Eq.\ref{Dt} and Eq.\ref{ratio}, to obtain the distance information,
ancillary data are needed in terms of
a few percent measurement of the spatially resolved velocity
dispersion of the lens galaxy, the LOS mass fluctuation and the highly resolved imaging from space telescopes.
Therefore, as in Jee et al. (2016), we set the following criteria:
1) the AGN image separation should be $>1''$ to distinguish them; 2) the third brightest image should be bright enough, its i-band magnitude $m_i<21$;
3) the lens galaxy should be bright enough $m_i<22$; 4) quadruply imaged lenses which carry more information to break the
Source-Position Transformation (SPT), such that the uncertainty of the lens modelling process is comparable
with the time-delay measurements, leading to percent level distance measurements from individual lenses.

With these assumptions, there will be $\sim55$ high-quality lenses of the 400 lenses mentioned above
selected from the mock LSST catalog~\citep{OM10} that can give both $D_{\Delta t}$ and $D_L$
measurements. However, this number may vary due to the real distributions of the lens galaxies and quasars. Besides,
the telescope observation strategy also limit the estimate, for example, the spectroscopic follow-up may require brighter lenses,
for a shallower limit $m_i<21$ of the lens galaxies, the number would be only $\sim35$.
In the best case, both time-delay measurements and the lens modelling should achieve several percent precision.
We take $\sim5\%$ precision for the distances as in~\citep{Jee2016,Linder2011}.
We also consider $\sim10\%$ precision which is allowed by current techniques for comparison.
These lenses will be ``blind analysed" that can effectively
control the systematic errors which would bias the results. The two dimensional distributions of $z_L$ and $z_S$
can be found in Jee et al. (2016). We marginalize $z_L$ and get the distribution of $z_S$ in Fig.\ref{zs}. As one can see,
a large part of the corresponding quasars have high redshifts. $\sim35$ systems are with $z_S>2$.

To make a forecast, we take a fiducial flat $\Lambda$CDM model with $H_0=70km/s/Mpc$, $\Omega_M=0.3$ for the simulation.
For each lensing system, given $z_L$ and $z_S$, we firstly calculate the fiducial values of $D_{\Delta t}$ and $D_L$,
then randomly generate 10000 realizations for each of them. The noise levels follow Gaussian distributions with uncertainties $5\%$ and $10\%$, respectively.
The uncertainties of $D_{\Delta t}$ and $D_L$ primarily come from the external
convergence and the velocity dispersion of the lens, respectively.
However, since measurements of $D_{\Delta t}$ and $D_L$ are based on the same lens model, one should consider the correlation between them unless one of the
distances have much larger uncertainty. According to the simulation by~\citep{Yildirim2019}, the correlation is positive.
We therefore try different correlation amplitudes for $D_{\Delta t}$ and $D_L$ with correlation coefficients $\rho=0.1, 0.4, 0.7$, respectively.

Then for each simulated $D_{\Delta t}$ and $D_L$ pair, we calculate
$D_S$ based on Eq.\ref{Ds}. At last, we get the distribution of $D_S$ along with its median value plus $16^{th}$ and $84^{th}$ percentiles.
Fig.\ref{example1} and Fig.\ref{example2} correspond to a typical case where $z_L=0.7,z_S=2.5$ for uncertainties $5\%$ and $10\%$ of $D_{\Delta t}$ and $D_L$, respectively.
We plot the simulated $D_{\Delta t}$ and $D_L$ distributions with different correlations
at the upper panels and calculate the corresponding Probability Density Distribution (PDF) of $D_S$ at the bottom panels.
One can see the constraint becomes tighter when the correlation is larger.
To show the dependence on the redshifts of the lens and source, the uncertainties of $D_{\Delta t}$ and $D_L$, and the correlation amplitude for the whole samples,
we plot Fig.~\ref{forecast} where $\sigma_{D_S}=(84^{th}percentile-16^{th}percentile)/2$ is taken as an estimate of the uncertainty.

\section{Summary and discussions}
In this work, we find another powerful cosmological application of strong gravitational lensing.
We propose to measure the angular diameter distances to quasars at high redshifts with strong lensing and apply the method to SDSS 1206+4332.
We also explore the power in the future LSST era and give the constraint dependence on the properties of the systems.
Rather than constraining a specific cosmological model, distances measured to the sources would benefit reconstructing the nature of the
Universe model-independently and directly at high redshifts. A further work will emulate the reconstruction.
Note that the measured high-redshift angular diameter distances can be used to find their maximum value and the corresponding
redshift since unlike luminosity distance, angular diameter distance would decrease if the redshift is larger than certain value $z\sim1.6$~\citep{Salzano2015}.

Very recently, Y${\i}$ld${\i}$r${\i}$m et al. (2019) presented a joint strong lensing and stellar dynamical framework
for future time-delay cosmography purposes. With the observations of high signal-to-noise integral field unit (IFU)
from the next generation of telescopes, they proved that $D_{\Delta t}$ can be constrained with $2.3\%$ uncertainty
and $D_L$ with $1.8\%$ at best for a system like RXJ1131. In such cases, we can acquire much more precise $D_S$ as well, making this idea
very promising. Note that RXJ1131 is the best case whereas an ordinary lens system
would give larger uncertainties.

Our method relies on the inputs of $D_{\Delta t}$ and $D_L$ measurements by the H0LiCOW-like lensing teams.
With more and more precise measurements, the intrinsic (unknown) systematic errors would be
quite important. If the $D_{\Delta t}$ and $D_L$ are biased, the inferred $D_S$ would also be biased.
The H0LiCOW team has adopted a blind analysis to control systematics. Data challenges, e.g., the Time Delay
Challenge~\citep{Liao2015} and the Lens Modelling Challenge~\citep{Ding2018} would reveal the systematics by algorithms. The systematics from
unknown physical process would be further revealed by independent approaches. Considering the $5\%$ and
$10\%$ uncertainties assumed in this work, a small systematic error, for example, $2\%$ would not bias the results.
There are great concerns about the lens modelling systematics being dominated by systematics~\citep{Schneider2013,Birrer2016,Tie2017}.
However, the point is no benefit from combining
lenses to constrain the cosmological models. In this work, we only focus on determining individual $D_S$.

\section{acknowledgments}
I thank the anonymous referee for his/her efforts to improve the quality of the paper,
and Simon Birrer for introducing the data of SDSS 1206+4332 on the H0LiCOW website.
This work was supported by the National Natural Science Foundation of China (NSFC) No. 11603015
and the Fundamental Research Funds for the Central Universities (WUT:2018IB012).

\clearpage


\begin{thebibliography}{}
\bibitem[Anselmi et al.(2019)]{Anselmi2019} Anselmi S., Corasaniti P.-S., Sanchez A. G., et al., 2019, PhRvD, 99, 123515
\bibitem[Arendse et al.(2019)]{Arendse2019} Arendse N., Agnello A., Wojtak R., 2019, arXiv: 1905.12000
\bibitem[Birrer et al.(2016)]{Birrer2016} Birrer S., Amara A., Refregier A., 2016, JCAP, 08, 020
\bibitem[Birrer et al.(2019)]{Birrer2019} Birrer S., Treu T., Rusu C. E., et al., 2019, MNRAS, 484, 4726
\bibitem[Betoule et al.(2014)]{Betoule2014} Betoule M., Kessler R., Guy J., et al., 2014, A\&A, 568, A22
\bibitem[Bernal et al.(2016)]{Bernal2016} Bernal J. L., Verde L., Riess A. G., 2016, JCAP, 10, 019
\bibitem[Chen \& Ratra(2012)]{Chen2012} Chen Y., Ratra B., 2012, A\&A, 543, A104
\bibitem[Chen et al.(2017)]{Chen2017} Chen Y., Kumar S., Ratra B., 2017, ApJ, 835, 86
\bibitem[Cai \& Yang(2017)]{Cai2017} Cai R.-G., Yang T., 2017, PhRvD, 95, 044024
\bibitem[Clarkson \& Zunckel(2010)]{Clarkson2010} Clarkson C., Zunckel C., 2010, PhRvL, 104, 211301
\bibitem[Courbin et al.(2018)]{Courbin2018} Courbin F., Bonvin V., Buckley-Geer E., et al., 2018,  A\&A, 609, A71
\bibitem[Denissenya et al.(2018)]{Denissenya2018} Denissenya M., Linder E. V., Shafieloo A., 2018, JCAP, 03, 041
\bibitem[Ding et al.(2015)]{Ding2015} Ding X., Biesiada M., Cao S., Li Z., Zhu Z.-H., 2015, ApJ, 803, L22
\bibitem[Ding et al.(2018)]{Ding2018} Ding X., Treu T., Shajib A. J., et al., 2018, arXiv: 1801.01506
\bibitem[Frieman et al.(2008)]{Frieman2008} Frieman J. A., Turner M. S., Huterer D., 2008, Annu. Rev. Astro. Astrophys., 46, 385
\bibitem[Freedman(2017)]{Freedman2017} Freedman W. L., 2017, Nature Astronomy, 1, 0121
\bibitem[Freedman et al.(2019)]{Freedman2019} Freedman W. L., Madore B. F., Hatt D., et al., 2019, arXiv:1907.05922
\bibitem[Falco et al.(1985)]{Falco1985} Falco E. E., Gorenstein M. V., Shapiro I. I., 1985, ApJ, 289, L1
\bibitem[Izzo et al.(2009)]{Izzo2009} Izzo L., Capozziello S., Govone G., Capaccioli M., 2009, A\&A, 508, 63
\bibitem[Jee et al.(2015)]{Jee2015} Jee I., Komatsu E., Suyu S. H., 2015, JCAP, 11, 033
\bibitem[Jee et al.(2016)]{Jee2016} Jee I., Komatsu E., Suyu S. H., Huterer D., 2016, JCAP, 04, 031
\bibitem[Jiang \& Kochanek(2007)]{Jiang2007} Jiang G., Kochanek C. S., 2007, ApJ, 671, 1568
\bibitem[Li et al.(2019)]{Li2019} Li E.-K., Du M., Xu L., 2019, arXiv: 1903.11433
\bibitem[Liao et al.(2015)]{Liao2015} Liao K., Treu T., Marshall P., et al., 2015, ApJ, 800, 11
\bibitem[Liao et al.(2018)]{Liao2018} Liao K., Ding X., Biesiada M., Fan X.-L., Zhu Z.-H., 2018, ApJ, 867, 69
\bibitem[Liao(2019)]{Liao2019} Liao K., 2019, ApJ, 871, 113
\bibitem[Linder(2011)]{Linder2011} Linder E. V., 2011, PhRvD, 84, 123529.
\bibitem[L'Huillier et al.(2019)]{LHuillier2019} L'Huillier B., Shafieloo A., Linder E. V., Kim A. G., 2019, MNRAS, 485, 2783
\bibitem[Moore(1994)]{Moore1994} Moore B., 1994, Nature, 370, 629
\bibitem[Moore(1999)]{Moore1999} Moore B., Ghigna S., Governato F., et al., 1999, ApJ, 524, L19
\bibitem[Oguri et al.(2005)]{Oguri2005} Oguri M., Inada N., Hennawi J. F., et al., 2005, ApJ, 622, 106
\bibitem[Oguri \& Marshall(2010)]{OM10} Oguri M., Marshall P. J., 2010, MNRAS, 405, 2579
\bibitem[Planck Collaboration(2018)]{Planck2018} Planck Collaboration, 2018, arXiv: 1807.06209
\bibitem[Paraficz \& Hjorth(2009)]{Paraficz2009} Paraficz D., Hjorth J., 2009, A\&A, 507, L49
\bibitem[Refsdal(1964)]{Refsdal1964} Refsdal S., 1964, MNRAS, 128, 307
\bibitem[Riess et al.(2019)]{Riess2019} Riess A. G., Casertano S., Yuan W., Macri L. M., Scolnic D., 2019, ApJ, 876, 85
\bibitem[Rusu et al.(2017)]{Rusu2017} Rusu C. E., Fassnacht C. D., Sluse D., et al., 2017, MNRAS, 467, 4220
\bibitem[Risaliti \& Lusso(2019)]{Risaliti2019} Risaliti G., Lusso E., 2019, Nature Astronomy, 3, 272
\bibitem[Schaefer(2003)]{Schaefer2003} Schaefer B. E., 2003, ApJ, 583, L67
\bibitem[Shafieloo et al.(2006)]{Shafieloo2006} Shafieloo A., Alam U., Sahni V., Starobinsky A. A., 2006, MNRAS, 366, 1081
\bibitem[Shafieloo(2007)]{Shafieloo2007} Shafieloo A., 2007, MNRAS, 380, 1573
\bibitem[Shafieloo et al.(2012)]{Shafieloo2012} Shafieloo A., Kim A. G., Linder E. V., 2012, PhRvD, 85, 123530
\bibitem[Seikel et al.(2012)]{Seikel2012} Seikel M., Clarkson C., Smith M., 2012, JCAP, 06, 026
\bibitem[Schutz(1986)]{Schutz1986} Schutz B. F., 1986, Nature, 323, 310
\bibitem[Shajib et al.(2018a)]{Shajib2018a} Shajib A. J., Treu T., Agnello A., 2018a, MNRAS, 473, 210
\bibitem[Shajib et al.(2018b)]{Shajib2018} Shajib A. J., Birrer S., Treu T., et al., 2018b, MNRAS, 483, 5649
\bibitem[Suyu et al.(2010)]{Suyu2010} Suyu S. H., Marshall P. J., Auger M. W., et al., 2010, ApJ, 711, 201
\bibitem[Suyu et al.(2017)]{Suyu2017} Suyu S. H., Bonvin V., Courbin F., et al., 2017, MNRAS, 468, 2590
\bibitem[Salzano et al.(2015)]{Salzano2015} Salzano V., Dabrowski M. P., Lazkoz R., 2015, PhRvL, 114, 101304
\bibitem[Schneier \& Sluse(2013)]{Schneider2013} Schneider P., Sluse D., 2013, A\&A, 559, A37
\bibitem[Treu(2010)]{Treu2010} Treu T., 2010, Annu. Rev. Astron. Astrophys. 48, 87
\bibitem[Treu \& Marshall(2016)]{Treu2016} Treu T., Marshall P. J., 2016, The Astronomy and Astrophysics Review, 24, 11
\bibitem[Taubenberger et al.(2019)]{Taubenberger2019} Taubenberger S., Suyu S. H., Komatsu E., 2019, arXiv:1905.12496
\bibitem[Tie \& Kochanek(2017)]{Tie2017} Tie S. S., Kochanek C. S., 2017, MNRAS, 473, 80
\bibitem[Wei(2010)]{Wei2010} Wei H., 2010, JCAP, 1008, 020
\bibitem[Wang et al.(2015)]{Wang2015} Wang F. Y., Dai Z. G., Liang E. W, 2015, New Astronomy Reviews, 67, 1
\bibitem[Wong et al.(2019)]{Wong2019} Wong K. C., Suyu S. H., Chen G. C.-F., et al., 2019, arXiv:1907.04869
\bibitem[Yang et al.(2015)]{Yang2015} Yang T., Guo Z.-K., Cai R.-G., 2015, PhRvD, 91, 123533
\bibitem[Y{\i}ld{\i}r{\i}m et al.(2019)]{Yildirim2019} Y{\i}ld{\i}r{\i}m A., Suyu S. H., Halkola A., 2019, arXiv:1904.07237
\bibitem[Zhao \& Wen(2018)]{Zhao2018} Zhao W., Wen L., 2018, PhRvD, 97, 064031


















\end{thebibliography}
\end{document}